\documentclass[sn-mathphys,Numbered]{sn-jnl}

\usepackage{multirow}%
\usepackage{amsmath,amssymb,amsfonts}%
\usepackage{amsthm}%
\usepackage{mathrsfs}%
\usepackage[title]{appendix}%
\usepackage{xcolor}%
\usepackage{textcomp}%
\usepackage{manyfoot}%
\usepackage{booktabs}%
\usepackage{algorithmicx}%
\usepackage{algpseudocode}%
\usepackage{listings}%

\usepackage{complexity}
\usepackage[shortlabels]{enumitem}
\usepackage{graphicx}
\usepackage[T1]{fontenc}
\usepackage[ruled]{algorithm2e}

\SetAlFnt{\small}
\SetAlCapFnt{\small}
\SetAlCapNameFnt{\small}
\SetAlCapHSkip{0pt}
\IncMargin{-\parindent}
\usepackage{amsthm}

\usepackage{graphics}
\usepackage{epsfig,verbatim}
\usepackage{arcs}
\usepackage{fontenc}
\usepackage{todonotes}
\usepackage{alltt}
\usepackage{amssymb,amsmath,amsfonts,epsf,url}
\usepackage{graphicx}
\usepackage{pstricks,pstricks-add,xcolor}
\usepackage{pst-node}
\usepackage{pst-coil}
\usepackage{tikz}
\usetikzlibrary{arrows,positioning,shapes,fit,calc,snakes,decorations.pathmorphing}
\tikzset{snake it/.style={decorate, decoration=snake}}

\pgfdeclarelayer{background}
\pgfsetlayers{background,main}

\usepackage{subfloat}

\theoremstyle{plain}
\newtheorem{theorem}{Theorem}
\newtheorem{lemma}{Lemma}%
\newtheorem{claim}{Claim}%
\newtheorem{definition}{Definition}%

\newcommand{\Oh}{\mathcal{O}}
\newcommand{\Ostar}{\mathcal{O^*}}

\newcommand{\nat}{\mathbb{N}}

\newcommand{\alg}{{\rm\bf FlipDist}}
\newcommand{\algo}{{\rm\bf FlipDist-I}}
\newcommand{\algos}{{\rm\bf FlipDist-S}}

\newcommand{\paramproblem}[4]{\noindent {\sc #1}
\\
{\bf Given:} #2\\
{\bf Parameter:} #3\\
{\bf Question:} #4}

\def\dist{{\rm Dist}}
\def\common{{\rm C}}

\newlength{\alginputwidth}
\newlength{\algboxwidth}

\newcommand{\algtitle}[1]{\underline{{\bf #1}} \vspace*{1mm}\\}

\newsavebox{\algbox}
\newsavebox{\captionboxx}
\newenvironment{algorithmnew}[2]%
    {
        \setlength{\algboxwidth}{\columnwidth}
        \addtolength{\algboxwidth}{-\columnsep}
        \addtolength{\algboxwidth}{-1mm}
        \setlength{\alginputwidth}{\algboxwidth}
        \addtolength{\alginputwidth}{-1.7cm}
        \begin{figure}[htbp]
            \vspace*{-1mm}
            \centering
            \begin{lrbox}{\captionboxx}
                \begin{minipage}[b]{\algboxwidth}
                    \centering
                    \caption{#1}
                    \label{#2}
                \end{minipage}
            \end{lrbox}
            \begin{lrbox}{\algbox}
                \begin{minipage}[b]{\algboxwidth}
                    \footnotesize
                    \vspace*{2mm}
    } 
    {
                    \vspace*{0.2mm}
               \end{minipage}
            \end{lrbox}
            \fbox{\usebox{\algbox}\hspace*{1mm}}
            \usebox{\captionboxx}
            \vspace*{-1mm}
      \end{figure}
    }
\newsavebox{\algcodebox}
\newenvironment{codeblock}%
    {
        \begin{enumerate}
            \setlength{\itemsep}{2pt}
            \setlength{\parsep}{0pt}
            \setlength{\topsep}{0pt}
            \setlength{\parskip}{0pt}
            \setlength{\partopsep}{0pt}
    } 
    {\end{enumerate}}
\newcommand{\step}{\item}

\begin{document}
\title{An $\Oh(3.82^{k})$ Time $\FPT$ Algorithm for Convex Flip Distance}  

\author[1]{\fnm{Haohong} \sur{Li}}\email{lih@lafayette.edu}
\author*[1]{\fnm{Ge} \sur{Xia}}\email{xiag@lafayette.edu}

\affil[1]{\orgdiv{Department of Computer Science}, \orgname{Lafayette College}, \orgaddress{\city{Easton}, \postcode{18040}, \state{PA}, \country{USA}}}

\keywords{Flip distance, Rotation distance, Triangulations, Exact algorithms, Parameterized complexity} 





\abstract{

Let ${\cal P}$ be a convex polygon in the plane, and let ${\cal T}$ be a triangulation of ${\cal P}$. An edge $e$ in ${\cal T}$ is called a diagonal if it is shared by two triangles in ${\cal T}$. A {\em flip} of a diagonal $e$ is the operation of removing $e$ and adding the opposite diagonal of the resulting quadrilateral to obtain a new triangulation of ${\cal P}$ from ${\cal T}$. The {\em flip distance} between two triangulations of ${\cal P}$ is the minimum number of flips needed to transform one triangulation into the other. The {\sc Convex Flip Distance} problem asks if the flip distance between two given triangulations of ${\cal P}$ is at most $k$, for some given parameter $k \in \nat$. It has been an important open problem to decide whether {\sc Convex Flip Distance} is \NP-hard. In this paper, we present an $\FPT$ algorithm for the {\sc Convex Flip Distance} problem that runs in time $\Oh(3.82^{k})$ and uses polynomial space, where $k$ is the number of flips. This algorithm significantly improves the previous best $\FPT$ algorithms for the problem. 

}
\maketitle


\section{Introduction} \label{sec:intro}

Let ${\cal P}$ be a convex polygon in the plane. A \emph{triangulation} of ${\cal P}$  adds edges between non-adjacent points in ${\cal P}$ to produce a planar graph ${\cal T}$ such that all faces except the outer face in ${\cal T}$ are triangles. The edges in the triangulation ${\cal T}$ not on the convex hull are called \emph{diagonals}. The number of diagonals in ${\cal T}$ is denoted by $\phi({\cal T})$.

A \emph{flip} of a diagonal $e$ in ${\cal T}$ removes $e$ and adds the opposite diagonal of the resulting quadrilateral, thus transforming ${\cal T}$ into another triangulation ${\cal T}'$ of ${\cal P}$. 
Given two triangulations ${\cal T}_{init}$ and ${\cal T}_{final}$ of ${\cal P}$, the \emph{flip distance} between them, denoted as $\dist({\cal T}_{init}, {\cal T}_{final})$, is the minimum number of flips required to transform ${\cal T}_{init}$ into ${\cal T}_{final}$ (or equivalently from ${\cal T}_{final}$ to ${\cal T}_{init}$). The {\sc Convex Flip Distance} problem is formally defined as: \\

\paramproblem{{\sc Convex Flip Distance}}{Two triangulation ${\cal T}_{init}$ and ${\cal T}_{final}$ of a convex polygon ${\cal P}$ in the plane.}{$k$.}{Is $\dist({\cal T}_{init}, {\cal T}_{final})$ at most $k$?} \\

The number of ways to triangulate a convex $(n+2)$-gon is $C_n$, the $n$-th Catalan number. Consequently, there exists an isomorphism between triangulations of a convex polygon and a plethora of counting problems, such as binary trees, Dyck paths, etc~\cite{enumbook}. In particular, there exists a bijection between the set of triangulations of a convex $(n+2)$-gon and the set of full binary trees with $n$ internal nodes. Furthermore, flipping an edge in a convex polygon triangulation corresponds to rotating a node in a binary tree, which is an essential operation for maintaining balanced binary search trees. The \emph{rotation distance} between two binary trees with the same number of nodes is the minimum number of rotations needed to transform one binary tree into the other. Thus, finding the flip distance between two triangulations of a convex $(n+2)$-gon is equivalent to finding the rotation distance between two binary trees with $n$ internal nodes.

Determining the computational complexity of {\sc Convex Flip Distance} (equivalently the rotation distance problem between binary trees) has been an important open problem. Despite extensive research on this problem~\cite{culik_wood,lucas3,lucas4,lucas20,lucas22,sleator,li_approx,fordham,cleary_app,cleary,lucas,kelk_kernel}, it is still unknown whether it is \NP-hard or not. 

In 1982, Culik and Wood~\cite{culik_wood} first studied the rotation distance problem between binary trees and proved an upper bound of $2n-2$, where $n$ is the number of internal nodes of the trees. In 1988, Sleator, Tarjan, and Thurston~\cite{sleator} improved the upper bound to $2n-6$, which is proven to be tight for $n \geq 11$ by Pournin~\cite{pournin}. 

In 1998, Li and Zhang gave two polynomial-time approximation algorithms \cite{li_approx} for computing the flip distance of convex polygon triangulations. One algorithm has an approximation ratio of $2- \frac{2}{4(d-1)(d+6)+1}$ given that each vertex is an endpoint of at most $d$ diagonals. Another algorithm has an approximation ratio of $1.97$ provided that both triangulations do not contain internal triangles. Cleary and St.~John~\cite{cleary_app} gave a linear-time 2-approximation algorithm for rotation distance in 2009.

In 2009, Cleary and St.~John~\cite{cleary} gave a kernel of size $5k$ for {\sc Convex Flip Distance} and presented an $\Oh^*((5k)^{k})$-time fixed-parameter tractable ($\FPT$) algorithm based on this kernel (the $\Ostar$ notation suppresses polynomial factors in the input size). The upper bound on the kernel size of {\sc Convex Flip Distance} was subsequently improved to $2k$ by Lucas~\cite{lucas}, who also gave an $\Oh^*(k^{k})$-time $\FPT$ algorithm for this problem. Very recently, Bosch-Calvo and Kelk improved the kernel size to $(1+\epsilon)k$ for any $\epsilon > 0$ in 2021 \cite{kelk_kernel}, although their result does not lead to an improved approximation algorithm over \cite{li_approx} and \cite{cleary_app}. 

 
The generalized version of the  {\sc Convex Flip Distance} problem, referred to as the {\sc General Flip Distance} problem, is the flip distance between triangulations of a point set in general positions in the plane.
The {\sc General Flip Distance} problem is also a fundamental and challenging problem, and has also been extensively studied~\cite{lawson,hurtadolower,mulzer,bosehurtado,cleary,hanke,urrutia,lubiw,pilz,kanjxiastacs,feng_improved}. 

In 1972, Lawson~\cite{lawson} gave an $\Oh(n^2)$ upper bound on {\sc General Flip Distance}, where $n$ is the number of points in ${\cal S}$~\cite{lawson}. The complexity of the {\sc General Flip Distance} problem was resolved in 2012 by Lubiw and Pathak~\cite{lubiw} who showed the problem to be $\NP$-complete. Simultaneously, and independently, the problem was shown to be $\APX$-hard by Pilz~\cite{pilz}. In 2015, Aichholzer Mulzer, and Pilz~\cite{mulzer} proved that the flip distance problem is $\NP$-complete for triangulations of a simple (but not convex) polygon.

Very recently, Kanj, Sedgwick, and Xia~\cite{fd17} presented the first $\FPT$ algorithm for {\sc General Flip Distance} that runs in $\Oh(n + k c^k)$, where $c \leq 2 \cdot 14^{11}$. Their approach defines a dependency relation for a sequence of flips: some flips required some other flips to be performed first. They proved that any topological sort of the directed acyclic graph (DAG) modeling this dependency relation yields the equivalent end result. Their algorithm simulates a ``non-deterministic walk'' that tries the possible flips to find a topological sort of the DAG representing an optimal solution. Refining this approach, Feng, Li, Meng, and Wang improved the $\FPT$ algorithm to run in $\Oh(n + k \cdot 32^k)$~\cite{feng_improved}, which currently stands as the best $\FPT$ algorithm for both {\sc General Flip Distance} and {\sc Convex Flip Distance}. No more efficient polynomial space algorithm was known for {\sc Convex Flip Distance} despite its structural properties. We note in passing here that although a straightforward algorithm based on breadth-first search (BFS) can find the flip distance between triangulations of a convex polygon in time $\Ostar(C_n) = \Ostar(4^n)$, where $C_n$ is the $n^{th}$ Catalan number, it requires exponential space. 

\section{Results}
In this paper, we present an $\FPT$ algorithm for the {\sc Convex Flip Distance} problem that runs in time $\Oh(3.82^k)$ and uses polynomial space, where $k$ is the number of flips. Instead of performing a ``non-deterministic walk'' as in~\cite{fd17,feng_improved}, our algorithm computes a topological sort of the DAG representing an optimal solution by repeatedly finding and removing
its source nodes. This approach allows us to take advantage of the structural properties of {\sc Convex Flip Distance} and design a simple yet significantly more efficient algorithm.

Our algorithm is closely related to algorithms in~\cite{fd17} and~\cite{feng_improved}. All three algorithms rely on the same structural results from~\cite{fd17} which state that (1) the flips in an optimal solution have a dependency relation that can be modeled as a DAG, and (2) any topological sort of this DAG yields an optimal solution. Therefore, three algorithms address the same problem: how to find a topological sort of this unknown DAG. The algorithm of Kanj, Sedgwick, and Xia~\cite{fd17} simulates a ``non-deterministic walk''. For any diagonal $e$, the algorithm will either flip $e$ (when this flip is a source in the DAG) or move on to one of $e$'s neighbors (when a neighbor of $e$ must be flipped before $e$ can be flipped). Through a sequence of such ``flip/move''-type local actions, a topological sort of the DAG (if exists) is found in time $\Oh(n + k c^k)$, where $c \leq 2 \cdot 14^{11}$. The algorithm of Feng, Li, Meng, and Wang~\cite{feng_improved}  refines the ``non-deterministic walk'' approach by reducing the number of the action types and streamlining the backtracking in the walk, resulting in an improved algorithm that runs in time $\Oh(n + k \cdot 32^k)$. Different from the previous ``walk''-based approach, our algorithm performs a topological sort of the DAG by repeatedly finding and removing source nodes in the DAG, similar to Kahn's algorithm~\cite{kahn}. After the current source nodes of the DAG are removed, the new source nodes can be found among the neighbors of the flipped diagonals. Based on this more efficient approach and by exploiting the structural properties of {\sc Convex Flip Distance}, we significantly improve the running time of our algorithm to $\Oh(3.82^k)$.

\section{Preliminaries}\label{sec:prelim}

\subsection{Flips, triangulations, and flip distance} 

For any flip $f$, we use the notation $f^\leftarrow$ to denote the underlying diagonal $e$ on which $f$ is performed, and the notation $f^\rightarrow$ to denote the new diagonal $\overline{e}$ added when $f$ is performed. For any two diagonals $e_1$ and $e_2$ in ${\cal T}$, we say they are {\em neighbors} if they appear in the same triangle and say they are {\em independent} if they are not neighbors. Note that two independent diagonals in ${\cal T}$ can share an endpoint, as long as they are not in the same triangle.

Let ${\cal T}$ and ${\cal T}'$ be two triangulations of ${\cal P}$. We refer to $({\cal T}, {\cal T}')$ as {\em a pair of triangulations} of ${\cal P}$. Denote by $\common({\cal T}, {\cal T}')$ the number of common diagonals shared by ${\cal T}$ and ${\cal T}'$. We say a sequence of flips $F=\langle f_1,\ldots, f_r \rangle$ {\em transforms} a triangulation ${\cal T}$ into ${\cal T}'$, denoted as ${\cal T} \xrightarrow{F} {\cal T}'$, if there exist triangulations ${\cal T}_0, \ldots, {\cal T}_r$ such that ${\cal T}_0 = {\cal T}$, ${\cal T}_r = {\cal T}'$, and performing flip $f_i$ in ${\cal T}_{i-1}$ results in ${\cal T}_i$, for $i=1, \ldots, r$. Such a transformation ${\cal T} \xrightarrow{F} {\cal T}'$  is referred to as a {\em path} from ${\cal T}$ to ${\cal T}'$ following $F$. The {\em length} of $F$ (equivalently, the length of the path ${\cal T} \xrightarrow{F} {\cal T}'$), denoted $|F|$, is the number of flips in it. The flip distance $\dist({\cal T}, {\cal T}')$ is the length of the shortest path between $T$ and $T'$. A sequence of flips $F$ such that ${\cal T} \xrightarrow{F} {\cal T}'$ is a shortest path is called an {\em optimal} (or {\em minimum}) {\em solution} of the pair $({\cal T}, {\cal T}')$. 


Let $f_i$ and $f_j$ be two flips in $F=\langle f_1,\ldots, f_r \rangle$ such that $1 \leq i < j\leq r$. The flip $f_j$ is said to be {\em adjacent} to the flip $f_i$, denoted $f_i \rightarrow f_j$, if $f_i^\rightarrow$ is a neighbor of $f_j^\leftarrow$ in ${\cal T}_{j-1}$. 
This adjacency relation defines a partial order among flips in $F$: if $f_i \rightarrow f_j$ then $f_i$ must precede $f_j$ because $f_i^\rightarrow$ is a neighbor of $f_j^\leftarrow$ at the moment when $f_j$ is performed. Therefore, the adjacency relation on the flips in $F$ can be naturally represented by a directed acyclic graph (DAG), denoted ${\cal D}_F$,
where the nodes of ${\cal D}_F$ are the flips in $F$, and its arcs represent the (directed) adjacencies in $F$.

Recall that a topological sort of a DAG is {\em any} ordering of its nodes that satisfies: For any directed arc $(u, v)$ in the DAG, $u$ appears before $v$ in the ordering. There could be many different topological sorts of ${\cal D}_F$, but the following lemma by Kanj, Sedgwick, and Xia~\cite{fd17} asserts that all of them yield the same outcome:

\begin{lemma}[\cite{fd17}] \label{lem:ts}
Let ${\cal T}_0$ be a triangulation and let $F=\langle f_1,\ldots, f_r \rangle$ be a sequence of flips such that ${\cal T}_0  \xrightarrow{F} {\cal T}_r$. Let $\pi(F)$ be a permutation of the flips in $F$ such that $\pi(F)$ is a topological sort of ${\cal D}_F$. Then $\pi(F)$ is a valid sequence of flips such that ${\cal T}_0 \xrightarrow{\pi(F)} {\cal T}_r$. Furthermore, the DAG ${\cal D}_{\pi(F)}$, defined based on the sequence $\pi(F)$, is the same directed graph as ${\cal D}_F$.
\end{lemma}


\begin{definition}\label{def:freeedge}
Let $({\cal T}_{init}, {\cal T}_{final})$ be a pair of triangulations. Let $e$ be a diagonal in ${\cal T}_{init}$. We say $e$ is a {\em free-diagonal with respect to ${\cal T}_{final}$} if flipping $e$ creates a new diagonal $\overline{e}$ that is in ${\cal T}_{final}$. When the context is clear, we simply refer to $e$ as a free-diagonal.
\end{definition}

\begin{lemma}\label{lem:freeedge-cross}
If $e_1$ and $e_2$ are two free-diagonals in ${\cal T}_{init}$, then $e_1$ and $e_2$ are independent.
\end{lemma}
\begin{proof}
Let $\overline{e}_1$ and $\overline{e}_2$ be the edges created by flipping $e_1$ and $e_2$ in ${\cal T}_{init}$, respectively. By Definition~\ref{def:freeedge}, both $\overline{e}_1$ and $\overline{e}_2$ are in ${\cal T}_{final}$. If  $e_1$ and $e_2$ are  neighbors, then $\overline{e}_1$ and $\overline{e}_2$ intersect each other, contradicting the fact that both $\overline{e}_1$ and $\overline{e}_2$ are in ${\cal T}_{final}$.
\end{proof}

The following lemma by Sleator, Tarjan and Thurston~\cite{sleator} shows that free-diagonals can be safely flipped and common diagonals will never be flipped in any shortest path.

\begin{lemma}[\cite{sleator}]\label{lem:path82}
Let $({\cal T}_{init}, {\cal T}_{final})$ be a pair of triangulations. (a) If ${\cal T}_{init}$ contains a free-diagonal $e$, then there exists a shortest path from ${\cal T}_{init}$ to ${\cal T}_{final}$ where $e$ is flipped first. (b) If ${\cal T}_{init}$ and ${\cal T}_{final}$ share a diagonal $e$ in common, then every shortest path from ${\cal T}_{init}$ to ${\cal T}_{final}$ never flips $e$.
\end{lemma}

In a sequence of flips, a previously non-free diagonal can become a free-diagonal only when one of its neighbors is flipped.

\begin{lemma}\label{lem:new-free}
Let $F=\langle f_1,\ldots, f_r \rangle$ be a sequence of flips such that ${\cal T}_{init} = {\cal T}_0  \xrightarrow{F} {\cal T}_r = {\cal T}_{final}$ is a shortest path. Let $e$ be a common diagonal of ${\cal T}_{i-1}$ and ${\cal T}_i$, for $1 \leq i \leq r$. If $e$ is not a free-diagonal in ${\cal T}_{i-1}$   and is a free-diagonal in ${\cal T}_i$, then $f_i^{\rightarrow}$ is a neighbor of $e$.
\end{lemma}
\begin{proof}
Suppose that $f_i^{\rightarrow}$ is not a neighbor of $e$. Then $Q_e$, the quadrilateral associated with $e$, remains the same in ${\cal T}_i$ as in ${\cal T}_{i-1}$. Therefore, flipping $e$ in ${\cal T}_{i-1}$ creates the same diagonal as flipping $e$ in ${\cal T}_i$, a contradiction to the fact that $e$ is not a free-diagonal in ${\cal T}_{i-1}$   and is a free-diagonal in ${\cal T}_i$.
\end{proof}

\begin{definition}\label{def:trivial}
A pair of triangulations $({\cal T}_{init}, {\cal T}_{final})$ of a convex polygon ${\cal P}$ is called {\em trivial} if $\dist({\cal T}_{init}, {\cal T}_{final}) = \phi({\cal T}_{init})-\common({\cal T}_{init}, {\cal T}_{final})$.

\end{definition}

\begin{lemma}\label{lem:trivial-linear}
If a pair of triangulations $({\cal T}_{init}, {\cal T}_{final})$ is trivial, then every flip in an optimal solution is a flip of a free-diagonal. Furthermore, it takes linear time to decide if a pair $({\cal T}_{init}, {\cal T}_{final})$ is trivial.
\end{lemma}
\begin{proof}
If $\dist({\cal T}_{init}, {\cal T}_{final}) = \phi({\cal T}_{init})-\common({\cal T}_{init}, {\cal T}_{final})$, then every flip in an optimal solution $F$ of $({\cal T}_{init}, {\cal T}_{final})$ must create an additional common diagonal between ${\cal T}_{init}$ and ${\cal T}_{final}$, which means that every flip in $F$ is performed on a free-diagonal.

To decide if a pair $({\cal T}_{init}, {\cal T}_{final})$ is trivial, first find all initial free-diagonals in ${\cal T}_{init}$ and add them to a queue. Then flip the free-diagonals in the queue. By Lemma~\ref{lem:new-free}, new free-diagonals must be neighbors of previous flips and hence can be found and added to the queue as the previous free-diagonals are flipped. The pair $({\cal T}_{init}, {\cal T}_{final})$ is trivial if and only if there are always free-diagonals in the queue to be flipped until ${\cal T}_{init}$ is transformed into ${\cal T}_{final}$. Since the flip distance between ${\cal T}_{init}$ and ${\cal T}_{final}$ is at most $2n-4$, where $n$ is the number of diagonals in ${\cal T}_{init}$ and ${\cal T}_{final}$~\cite{pournin}, this process takes linear time. 
\end{proof}

\begin{definition}\label{def:safeset}
Let $({\cal T}_{init}, {\cal T}_{final})$ be a pair of triangulations. Let $I$ be a set of diagonals in ${\cal T}_{init}$. We say $I$ is a {\em safe-set of diagonals with respect to ${\cal T}_{final}$}, or simply {\em safe-set}, if 
\begin{enumerate}
    \item the diagonals in $I$ are pair-wise independent, and 
    \item for any permutation $\pi(I)$ of $I$, there is a shortest path from ${\cal T}_{init}$ to ${\cal T}_{final}$ such that the diagonals in $I$ are flipped first, in the same order as $\pi(I)$.
\end{enumerate}
\end{definition}

The set of diagonals corresponding to the source nodes of ${\cal D}_F$ is a safe-set.

\begin{lemma}\label{lem:source}
Let $F=\langle f_1,\ldots, f_r \rangle$ be a sequence of flips such that ${\cal T}_{init} = {\cal T}_0  \xrightarrow{F} {\cal T}_r = {\cal T}_{final}$ is a shortest path. Let $SC =\{ f_{sc_1}, \ldots, f_{sc_{l}} \}$ be the set of source nodes in ${\cal D}_F$. The set of diagonals $I = \{f_{sc_1}^{\leftarrow}, \ldots, f_{sc_{l}}^{\leftarrow} \}$ is a safe-set in ${\cal T}_{init}$. 
\end{lemma}
\begin{proof}
Let $f_{sc_i}, f_{sc_j} \in SC$, $i \neq j$, be two source nodes in ${\cal D}_F$. By Lemma~\ref{lem:ts}, we may assume that $f_{sc_i}$ is the first flip in $F$. If $f_{sc_i}^{\leftarrow}$ and $f_{sc_j}^{\leftarrow}$ share a triangle in ${\cal T}_{init}$, after flipping $f_{sc_i}$, $f_{sc_i}^{\rightarrow}$ and $f_{sc_j}^{\leftarrow}$ share a triangle in ${\cal T}_1$, and hence by Lemma 3.10 of \cite{fd17}, there is a directed path from $f_{sc_i}$ to $f_{sc_j}$ in ${\cal D}_F$, contradicting to the fact that $f_{sc_i}$ is a source in ${\cal D}_F$. Therefore, the diagonals in $I$ are independent.

For an arbitrary permutation of $\pi(I)$, there is a topological sort of ${\cal D}_F$ that begins with the flips in $SC$ according to the order of $\pi(I)$. By  Lemma~\ref{lem:ts}, there is a shortest path between  ${\cal T}_{init}$ and ${\cal T}_{final}$ that flips the diagonal in $I$ first, in the order of $\pi(I)$. 
Therefore, $I$ is a safe-set.
\end{proof}

\subsection{Counting matchings in binary trees}
The analysis of the running time of our algorithm uses the following lemma regarding the number of matchings in a binary tree, which may be of independent interest. Note that the inequalities in the lemma are tight when the tree is a path.

\begin{lemma}\label{lem:matching}
Let $T_u$ be a binary tree rooted at a node $u$ with $n$ nodes and $n-1$ edges. A matching in $T_u$ is a subset of edges in $T_u$ that do not share any endpoint (we consider an empty set to be a matching). Let ${\cal E}_u$ be the set of matchings in $T_u$. Let ${\cal E}_u^-$ be the set of matchings in $T_u$ that exclude the edge(s) incident on $u$. Then $|{\cal E}_u^-| \leq F_{n}$ and $|{\cal E}_u| \leq F_{n+1}$, where $F_n$ is the $n$-th Fibonacci number.
\end{lemma}
\begin{proof}
Let ${\cal E}_u^+$ be the set of matchings in $T_u$ that include exactly one edge incident on $u$. Then $|{\cal E}_u| = |{\cal E}_u^+| + |{\cal E}_u^-|$. The lemma is proven by induction on $n$. 

When $n = 1$, $T_u$ is a leaf. Thus ${\cal E}_u^+ = \emptyset$ and ${\cal E}_u^- = \{\emptyset\}$. Therefore, $|{\cal E}_u^+| = 0 = F_0$, $|{\cal E}_u^-| = 1 = F_1$, and  $|{\cal E}_u| = |{\cal E}_u^+| + |{\cal E}_u^-| = 1 = F_2$.

When $n = 2$, $T_u$ has a single edge that connects $u$ to a single leaf child $v$. Thus, ${\cal E}_u^+ = \{uv\}$ and ${\cal E}_u^- = \{\emptyset\}$. Therefore, $|{\cal E}_u^+| = 1 = F_1$, $|{\cal E}_u^-| = 1 = F_2$, and  $|{\cal E}_u| = |{\cal E}_u^+| + |{\cal E}_u^-| = 2 = F_3$.

Now suppose that $n \geq 3$. We distinguish 2 cases.
\begin{enumerate}[(a)]
    \item $u$ has only one child $v$. In this case, $T_v$, the subtree rooted at $v$, has $n-1$ nodes and $n-2$ edges. Every matching in ${\cal E}_u^+$ includes $uv$ and hence excludes edges in $T_v$ incident on $v$. Thus $|{\cal E}_u^+| = |{\cal E}_v^-| \leq F_{n-1}$ by the induction hypothesis. Every matching in ${\cal E}_u^-$ excludes $uv$ and hence is also a matching in $T_v$. Therefore, $|{\cal E}_u^-| = |{\cal E}_v| \leq F_{(n-1)+1} = F_n$ by the induction hypothesis. Finally, $|{\cal E}_u| = |{\cal E}_u^+| + |{\cal E}_u^-| \leq F_{n-1} + F_{n} = F_{n+1}$. The statement is true.

    \item $u$ has two children $v,w$. Let the number of nodes in $T_v$ and $T_w$ be $n_1$ and $n_2$, respectively. Then $n = n_1+n_2+1$. We consider ${\cal E}_u^-$ and ${\cal E}_u^+$ separately:

\begin{enumerate}[(i)]
    \item The matchings in ${\cal E}_u^-$ exclude both $uv$ and $uw$. Therefore, $|{\cal E}_u^-| = |{\cal E}_v|\cdot|{\cal E}_w| \leq F_{n_1+1} F_{n_2+1} \leq F_{n_1+n_2+1} \leq F_{n}$ by the induction hypothesis and the Honsberger's Identity of Fibonacci numbers.
    \item The matchings in ${\cal E}_u^+$ include exactly one of the edges $uv$ and $uw$. The number of matchings that include $uv$ and exclude $uw$ is $|{\cal E}_v^-|\cdot|{\cal E}_w| \leq F_{n_1} F_{n_2+1}$ by the induction hypothesis. The number of matchings that include $uw$ and exclude $uv$ is $|{\cal E}_v|\cdot|{\cal E}_w^-| \leq F_{n_1+1} F_{n_2}$ by the induction hypothesis. Therefore, $|{\cal E}_u^+| \leq F_{n_1} F_{n_2+1} + F_{n_1+1} F_{n_2}$.
\end{enumerate}
Finally, $|{\cal E}_u| = |{\cal E}_u^-| + |{\cal E}_u^+|  \leq F_{n_1+1} F_{n_2+1} +F_{n_1} F_{n_2+1} + F_{n_1+1} F_{n_2} = F_{n_1+2} F_{n_2+1} + F_{n_1+1} F_{n_2} = F_{n_1+n_2+2} = F_{n+1}$, where the second last equality is the Honsberger's Identity of Fibonacci numbers.
\end{enumerate}

This completes the proof.
\end{proof}

\subsection{Parameterized complexity.}
A {\em parameterized problem} is a set of instances of the form
$(x, k)$, where $x$ is the input instance and $k \in \nat$
is the {\it parameter}. A parameterized problem is
{\it fixed-parameter tractable} ($\FPT$) if there
is an algorithm that solves the problem in time $f(k)|x|^{c}$, where $f$ is a computable function and $c > 0$ is
a constant. We refer to~\cite{fptbook,rolfbook} for more information about parameterized complexity.

\section{The algorithm and its analysis}\label{sec:algorithm}

\subsection{The basic ideas}

Given an instance $({\cal T}_{init}, {\cal T}_{final}, k)$ of {\sc Convex Flip Distance}, our algorithm decides whether there is a sequence of flips $F=\langle f_1,\ldots, f_r \rangle$, $r\leq k$ that transforms ${\cal T}_{init}$ into ${\cal T}_{final}$. 

By Lemma~\ref{lem:path82}, common diagonals will never be flipped and free diagonals can be safely flipped. Thus, we can assume that $({\cal T}_{init}, {\cal T}_{final})$ does not have any common diagonals, i.e., $\common({\cal T}_{init}, {\cal T}_{final}) = 0$, and that there are no free-diagonals in the initial triangulation ${\cal T}_{init}$. Therefore, we can assume $n \leq k \leq 2n-4$, where $n = \phi({\cal T}_{init})$~\cite{pournin}.

In Preliminaries, we showed that we can represent a minimum solution $F$ to an instance $({\cal T}_{init}, {\cal T}_{final})$ using a DAG ${\cal D}_F$, and that any topological sort of ${\cal D}_F$
is a minimum solution (Lemma~\ref{lem:ts}). Therefore, to solve an instance of {\sc Convex Flip Distance}, it suffices to compute a topological sort of ${\cal D}_F$.

Intuitively, our algorithm computes a topological sort of ${\cal D}_F$ by repeatedly finding and removing its source nodes, similar to Kahn's algorithm~\cite{kahn}. The algorithm uses a branch-and-bound approach to find and flip the source nodes in ${\cal D}_F$. 

This seemingly simple approach faces two technical challenges. The design of our algorithm revolves around addressing them. 

First, how to find the source nodes of the unknown ${\cal D}_F$ without trying all possible subsets of the diagonals? The set $I$ of initial source nodes of ${\cal D}_F$ must be a subset of independent diagonals in ${\cal T}_{init}$. Our algorithm enumerates all subsets of independent diagonals in ${\cal T}_{init}$, whose number is at most $F_{n+1}$, the $(n+1)$-th Fibonacci number (Lemma~\ref{lem:iterate}). Afterward, any new source node $f_j$ of ${\cal D}_F$ must be adjacent to a previous source node $f_i$. This means that the underlying diagonal of $f_j$ is a neighbor of the new diagonal created by $f_i$. Therefore, when flipping a source node, our algorithm adds all neighbors of the new diagonal to a candidate pool $S$ from which new source nodes are chosen. In fact, the neighbors of the new diagonal are added as {\em pairs} in $S$ because at most one diagonal in each pair may be chosen as a new source node. The next set of new source nodes of ${\cal D}_F$ will be chosen by branching on the edge-pairs in $S$. This method significantly reduces the search space of the source nodes.

Second, how to flip free-diagonals without increasing the branching factor of the algorithm? By Lemma~\ref{lem:path82}, any free-diagonals can be safely flipped. If there is a free-diagonal $e$ in ${\cal T}_{init}$, our algorithm flips it to create a new diagonal $\overline{e}$. Since $\overline{e}$ is a common diagonal and hence will never be flipped again, the instance $({\cal T}_{init}, {\cal T}_{final})$ is then partitioned along $\overline{e}$ into two smaller instances. The candidate pool $S$ is also partitioned accordingly and passed on to the two smaller instances. This method, through careful analysis, allows the free-diagonals to be flipped ``for free'' essentially.

\subsection{The algorithm}

\vspace*{0.5cm}
\renewcommand{\figurename}{Fig.}
\setcounter{figure}{0}

\begin{algorithmnew}{The function \alg.}{alg:wholealgo}
\small

\algtitle{\alg{\rm (${\cal T}_{init}$, ${\cal T}_{final}$, $k$)}}
     \textbf{Input:} Two triangulations ${\cal T}_{init}$ and ${\cal T}_{final}$, a parameter $k$ \\
     \textbf{Precondition:} $\common({\cal T}_{init}, {\cal T}_{final}) = 0$ and there is no free-diagonals in ${\cal T}_{init}$ \\
     \textbf{Output:} {\em True} if $\dist({\cal T}_{init}, {\cal T}_{final}) \leq k$; {\em False} otherwise.
\begin{codeblock}
\step[0.] If $\phi({\cal T}_{init}) > k$, return {\em False}.

\item[1.] Iterate through all subsets $I$ of independent diagonals in ${\cal T}$, as follows:
\begin{itemize}[leftmargin=0.35in]
\item[1.1.] For each diagonal $e \in {\cal T}$, if none of the neighbors of $e$ is in $I$, branch on two choices: (1) include $e$ in $I$; (2) do not include $e$ in $I$. 
\item[1.2.] At the end of the branching, if $I$ is non-empty, do: \\If \algo$({\cal T}_{init}, {\cal T}_{final}, k, I)$ returns {\em True}, then return {\em True}. 
\end{itemize}

\item[2.] Return {\em False}.

\end{codeblock}
\end{algorithmnew}

\begin{algorithmnew}{The function \algo.}{alg:algoi}
\small

\algtitle{\algo{\rm (${\cal T}_{init}$, ${\cal T}_{final}$, $k$, $I$)}}
     \textbf{Input:} Two triangulations ${\cal T}_{init}$ and ${\cal T}_{final}$, a parameter $k$, and a set of independent diagonals $I$, where $I \neq \emptyset$.    \\
     \textbf{Precondition:} $\common({\cal T}_{init}, {\cal T}_{final}) = 0$ and there are no free-diagonals in ${\cal T}_{init}$ \\
     \textbf{Output:} {\em True} if $\dist({\cal T}_{init}, {\cal T}_{final}) \leq k$; {\em False} otherwise.
\begin{codeblock}
\step[0.] If $\phi({\cal T}_{init}) > k-|I|$, return {\em False}; If $\phi({\cal T}_{init}) = 0$ and $k \geq 0$, return {\em True}.

\step[1.] Create an empty set $S$. For each edge $e$ in $I$, do:
\begin{itemize}[leftmargin=0.35in]
\item[1.1.] Flip $e$ to create a new edge $\overline{e}$.
\item[1.2.] Let $\Delta_1 = \{e_1,e_1',\overline{e}\}$ and $\Delta_2 = \{e_2, e_2',\overline{e}\}$ be the triangles on either side of $\overline{e}$. Add the pairs $(e_1,e_1')$ and $(e_2,e_2')$ to the set $S$.
\end{itemize}

\step[2.] Return \algos$(\overline{\cal T}_{init}, {\cal T}_{final}, k-|I|, S)$, where $\overline{\cal T}_{init}$ is the triangulation resulting from ${\cal T}_{init}$ after all edges in $I$ are flipped and $S = \{(e_1,e_1'), \ldots, (e_l, e_l')\}$ is the set of edge-pairs created in Step 1.

\end{codeblock}
\end{algorithmnew}

\renewcommand{\figurename}{Fig.}

\begin{algorithmnew}{The function \algos.}{alg:algos}
\small

\algtitle{\algos{\rm (${\cal T}_{init}$, ${\cal T}_{final}$, $k$, $S$)}}
     \textbf{Input:} Two triangulations ${\cal T}_{init}$ and ${\cal T}_{final}$, a parameter $k$, and a set of edge-pairs $S$.    \\
     \textbf{Precondition:} $\common({\cal T}_{init}, {\cal T}_{final}) = 0$ \\
     \textbf{Output:} {\em True} if $\dist({\cal T}_{init}, {\cal T}_{final}) \leq k$; {\em False} otherwise.
\begin{codeblock}
\step[0.] If $\phi({\cal T}_{init}) > k$, return {\em False}; If $\phi({\cal T}_{init}) = 0$ and $k \geq 0$, return {\em True}.

\step[1.] If there is a free-diagonal $e$ in ${\cal T}_{init}$, do:

\begin{itemize}[leftmargin=0.35in]

\item[1.1.] Remove any edge-pair containing $e$ from $S$. Flip the free-diagonal $e$ to create a new diagonal $\overline{e} \in {\cal T}_{final}$. Let $\overline{\cal T}_{init}$ be the triangulation after flipping $e$.

\item[1.2.] Let $\Delta_1 = \{e_1,e_1',\overline{e}\}$ and $\Delta_2 = \{e_2,e_2',\overline{e}\}$ be the triangles on either side of $\overline{e}$ in $\overline{\cal T}_{init}$. Add the edge-pairs $(e_1, e_1')$ and $(e_2, e_2')$ to $S$.

\item[1.3.] Partition the triangulations $\overline{\cal T}_{init}$ and ${\cal T}_{final}$ along $\overline{e}$ into $\{\overline{\cal T}_{init}^1,  \overline{\cal T}_{init}^2\}$ and $\{{\cal T}_{final}^1, {\cal T}_{final}^2\}$, respectively. Partition the edge-pairs $S$ into $\{S_1, S_2\}$ accordingly so that $S_1$ is in $\overline{\cal T}_{init}^1$  and $S_2$ is in $\overline{\cal T}_{init}^2$. Let $n_1=\phi(\overline{\cal T}_{init}^1)$ and $n_2 = \phi(\overline{\cal T}_{init}^2)$.

\item[1.4.] If $(\overline{\cal T}_{init}^1, {\cal T}_{final}^1)$ is trivial, return \algos$(\overline{\cal T}_{init}^2,~ {\cal T}_{final}^2,~ k - 1 - n_1,~ S_2)$.

\item[1.5.] If $(\overline{\cal T}_{init}^2, {\cal T}_{final}^2)$ is trivial, return \algos$(\overline{\cal T}_{init}^1,~ {\cal T}_{final}^1,~ k - 1 - n_2,~ S_1)$.

\item[1.6.] Find the smallest $k_1$ between $n_1+1$ and $k-2-n_2$ such that \\\algos$(\overline{\cal T}_{init}^1, {\cal T}_{final}^1, k_1, S_1)$ returns {\em True}; If such a $k_1$ does not exist, return {\em False}.

\item[1.7.] Return \algos$(\overline{\cal T}_{init}^2,~ {\cal T}_{final}^2,~ k - 1 - k_1,~ S_2)$.
\end{itemize}

\item[2.] Iterate through all non-empty subsets $I$ of independent diagonals in $\bigcup S$, as follows:
\begin{itemize}[leftmargin=0.35in]
\item[2.1.] For each edge pair $(e_i,e_i') \in S$, branch on up to three choices of their membership in $I$: (1) include neither $e_i$ nor $e_i'$; (2) include $e_i$ but not $e_i'$; and (3) include $e_i'$ but not $e_i$. Skip a choice if it tries to include a non-diagonal edge or a diagonal with a neighbor already included in $I$.
\item[2.2.] At the end of branching, if $I$ is non-empty, do:\\If \algo$({\cal T}_{init}, {\cal T}_{final}, k, I)$ returns {\em True}, then return {\em True}. 
\end{itemize}
\item[3.]  Return {\em False}.

\end{codeblock}
\end{algorithmnew}

\renewcommand{\figurename}{Fig.}

The algorithm's main function \alg~(Fig.~\ref{alg:wholealgo}) iterates through all subsets of independent diagonals in ${\cal T}$. For each non-empty subset $I$ of independent diagonals that represents the set of initial source nodes of a DAG ${\cal D}_F$, two mutually recursive functions \algo~(Fig.~\ref{alg:algoi}) and \algos~(Fig.~\ref{alg:algos}) are invoked to repeatedly remove the source nodes and find the set of new source nodes in ${\cal D}_F$. Along the way, whenever a free-diagonal is flipped to create a common diagonal, the instance is partitioned into smaller isolated instances.

Specifically, \algo(${\cal T}_{init}$, ${\cal T}_{final}$, $k$, $I$) performs flips on all diagonals in $I$. Suppose that a diagonal $e \in I$ is flipped to create a new edge $\overline{e}$. There are two triangles $\Delta_1 = \{e_1,e_1',\overline{e}\}$ and $\Delta_2 = \{e_2, e_2',\overline{e}\}$ on opposite sides of $\overline{e}$. The edges $e_1,e_1',e_2, e_2'$ are candidates of the new source nodes in ${\cal D}_F$. Since $(e_1,e_1')$ are neighbors and so are $(e_2,e_2')$, at most one edge can be chosen from each pair as a new source node. Therefore, the pairs $(e_1,e_1')$ and $(e_2,e_2')$ are added to a candidate pool $S$. After all diagonals in $I$ are flipped, the current triangulations are $\overline{\cal T}_{init}$ and ${\cal T}_{final}$, the current parameter is $k-|I|$ since $|I|$ flips are already performed, and the candidate poll is $S$, all of which are passed to \algos~as parameters. 

\algos(${\cal T}_{init}$, ${\cal T}_{final}$, $k$, $S$) first flips free-diagonals (if any) in ${\cal T}_{init}$. When a free-diagonal $e$ is flipped, a common diagonal $\overline{e}$ is created, and by Lemma~\ref{lem:path82}, the instance can be safely partitioned along $\overline{e}$ into two smaller isolated sub-instances, which can be solved recursively in a divide-and-conquer approach. In order to determine the parameters of the sub-instances, it is necessary to find the smallest parameter $k_1$ such that the first sub-instance returns {\em True}. The parameter of the second sub-instance is then set to be $k-k_1$. If there are no free-diagonals, \algos~branches on the pairs of edges in $S$ to form the safe-set $I$ for the next round. For each edge-pair $(e_i,e_i') \in S$ in $S$, $I$ may include neither, only $e_i$, or only $e_i'$, forming a three-way branching.

\subsection{Analysis of the algorithm}

In the following, we will prove the correctness of the algorithm and analyze its running time. We start by proving the following invariant for both \algo~and \algos:

\begin{claim}\label{cla:all}
For every new diagonal created in \algo~and \algos, its neighbors are contained in $\bigcup S$, which is the union of the edge-pairs in $S$.
\end{claim}
\begin{proof}
In \algo, after each diagonal $e \in I$ is flipped, the neighbors of the new diagonal $\overline{e}$ are included in the edge-pairs in $S$. Thus in \algo, $\bigcup S$ contains all neighbors of the new diagonals in $\overline{\cal T}_{init}$.

\algos$({\cal T}_{init}^i, {\cal T}_{final}^i, k, I)$~starts by flipping free-diagonals (if any) in ${\cal T}_{init}$. If there is a free-diagonal $e$ in ${\cal T}_{init}$, then by Lemma~\ref{lem:path82}, $e$ can be safely flipped to create a new diagonal $\overline{e} \in {\cal T}_{final}$. Before $e$ is flipped to create a new diagonal $\overline{e}$, any pair $(e, e')$ containing $e$ is removed from $S$. Let $\overline{\cal T}_{init}$ be the triangulation after flipping $e$. Let $\Delta_1 = \{e_1,e_1',\overline{e}\}$ and $\Delta_2 = \{e_2,e_2',\overline{e}\}$ be the triangles on either side of $\overline{e}$ in $\overline{\cal T}_{init}$. Two edge-pairs $(e_1, e_1')$ and $(e_2, e_2')$ are added to $S$. Note that if $(e, e')$ is a pair in $S$ before the free-diagonal $e$ is flipped, then $e'$ must be contained in one of two new pairs added to $S$ after $e$ is flipped. Therefore, after flipping $e$, $\bigcup S$ still contains all neighbors of the new diagonals in $\overline{\cal T}_{init}$, and hence the claim is true.
\end{proof}

\begin{lemma}\label{thm:correct}
Let $({\cal T}_{init}, {\cal T}_{final})$ be a pair of triangulations that do not share any common diagonal. Let $I$ be a set of independent diagonals in ${\cal T}_{init}$. Let $S$ be a set of edge-pairs in ${\cal T}_{init}$ and $\bigcup S$ be the union of the edge-pairs in $S$.
\begin{itemize}
\item[(a)] If there is a minimum solution $F$ of $({\cal T}_{init}, {\cal T}_{final})$ such that $I$ is the set of underlying diagonals of the source nodes of ${\cal D}_F$, then \algo$({\cal T}_{init}, {\cal T}_{final}, k, I)$ returns {\em True} if and only if $F$ has at most $k$ flips.
\item[(b)] If there is a minimum solution $F$ of $({\cal T}_{init}, {\cal T}_{final})$ such that the set of underlying diagonals of the source nodes of ${\cal D}_F$ is a subset of $\bigcup S$, then \algos$({\cal T}_{init}, {\cal T}_{final}, k, S)$ returns {\em True} if and only if $F$ has at most $k$ flips.
\end{itemize}

\end{lemma}
\begin{proof}
The proof is by mutual induction on $k$. For the base case when $k=0$, both functions will return true if and only if ${\cal T}_{init} = {\cal T}_{final}$. The statements are true.

Based on the inductive hypothesis, the inductive step is proven in two parts.
\begin{enumerate}[(a)]
    \item First consider \algo$({\cal T}_{init}, {\cal T}_{final}, k, I)$. By Lemma~\ref{lem:source}, $I$ is a safe-set and hence for any permutation $\pi(I)$ of $I$, there is a shortest path from ${\cal T}_{init}$ to  ${\cal T}_{final}$ such that the diagonals in $I$ are flipped first according to the order of $\pi(I)$. Therefore, $\dist({\cal T}_{init}, {\cal T}_{final}) \leq k$ if and only if $\dist(\overline{\cal T}_{init}, {\cal T}_{final}) \leq k - |I|$, where $\overline{\cal T}_{init}$ is the resulting triangulation after the diagonals in $I$ are flipped. After the set of diagonals $I$ corresponding to the source nodes of ${\cal D}_F$ are flipped and removed from ${\cal D}_F$, the set of diagonals corresponds to the new source nodes in ${\cal D}_F$ must be neighbors of the new diagonals in $\overline{\cal T}_{init}$ and hence by Claim~\ref{cla:all} is a subset of $\bigcup S$. Therefore, when \algo$({\cal T}_{init}, {\cal T}_{final}, k, I)$ calls \algos$(\overline{\cal T}_{init}, {\cal T}_{final}, k-|I|, S)$, by the inductive hypothesis, it returns {\em True} if and only if $\dist(\overline{\cal T}_{init}, {\cal T}_{final}) \leq k - |I|$, as required.

\item Now consider \algos$({\cal T}_{init}, {\cal T}_{final}, k, S)$ in two cases:
\begin{enumerate}[(i)]
\item If ${\cal T}_{init}$ has a free-diagonal $e$, then by Lemma~\ref{lem:path82}, there is a shortest path from ${\cal T}_{init}$ to  ${\cal T}_{final}$ such that $e$ is flipped first to create a new diagonal $\overline{e}$ that is shared by $\overline{\cal T}_{init}$ and ${\cal T}_{final}$. Again by Lemma~\ref{lem:path82}, no shortest paths from $\overline{\cal T}_{init}$ to ${\cal T}_{final}$ will flip $\overline{e}$. Therefore, the triangulations $\overline{\cal T}_{init}$ and ${\cal T}_{final}$ can be safely partitioned along $\overline{e}$ into $\{\overline{\cal T}_{init}^1,  \overline{\cal T}_{init}^2\}$ and $\{{\cal T}_{final}^1, {\cal T}_{final}^2\}$, respectively. By Claim~\ref{cla:all}, after $e$ is flipped, the set $S$ still contains all neighbors of the new diagonals, which is partitioned accordingly to $S_1$ and $S_2$. There is a path of length at most $k$ from ${\cal T}_{init}$ to ${\cal T}_{final}$ if and only if there is a path of length at most $k_1$ from ${\cal T}_{init}^1$ to ${\cal T}_{final}^1$ and a path of length at most $k_2$ from ${\cal T}_{init}^2$ to ${\cal T}_{final}^2$ such that $k_1+k_2 = k-1$. Furthermore, since both $(\overline{\cal T}_{init}^1,  \overline{\cal T}_{final}^1)$ and $({\cal T}_{init}^2, {\cal T}_{final}^2)$ can be assumed to be non-trivial after Steps 1.4 and 1.5, we have $k_1 \geq n_1+1$ and $k_2 \geq n_2+1$. Therefore the search range of $k_1$ is between $n_1+1$ and $k-1 - (n_2+1) = k-2-n_2$. It is easy to see that $S_1$ and $S_2$ satisfy the condition of Part (b) for $\overline{\cal T}_{init}^1$ and $\overline{\cal T}_{init}^2$, respectively. By the inductive hypothesis, \algos$(\overline{\cal T}_{init}^1, {\cal T}_{final}^1, k_1, S_1)$ returns {\em True} if and only if there is a path of length at most $k_1$ from ${\cal T}_{init}^1$ to ${\cal T}_{final}^1$, and \algos$(\overline{\cal T}_{init}^2,~ {\cal T}_{final}^2,~ k - 1 - k_1,~ S_2)$ returns {\em True} if and only if there is a path of length at most $k - 1 - k_1$ from $\overline{\cal T}_{init}^2$ to ${\cal T}_{final}^2$.
Therefore,  \algos$(\overline{\cal T}_{init}, {\cal T}_{final}, k-|I|, S)$ returns {\em True} if and only if there is a path of length at most $k$ from ${\cal T}_{init}$ to ${\cal T}_{final}$, as required.

\item If ${\cal T}_{init}$ has no free-diagonals, then \algos$({\cal T}_{init}, {\cal T}_{final}, k, S)$ enumerates all subsets $I$ of independent diagonals and calls \algo$({\cal T}_{init}, {\cal T}_{final}, k, I)$.
For each edge-pair $(e_1,e_2) \in S$, the algorithm branches on up to three possible choices: 1) include neither $e_1$ nor $e_2$, 2) include $e_1$ but not $e_2$, and 3) include $e_2$ but not $e_1$. Since $e_1$ and $e_2$ are neighbors, they cannot both belong to $I$. 
\algos$({\cal T}_{init}, {\cal T}_{final}, k, S)$ returns {\em True} if and only if there is an enumerated independent subset $I$ of $\bigcup S$ such that \algo$({\cal T}_{init}, {\cal T}_{final}, k, I)$ returns {\em True}. 
If $\dist({\cal T}_{init},{\cal T}_{final}) \leq k$, then by Claim~\ref{cla:all}, the set of the underlying diagonals of the source nodes of ${\cal D}_F$ is a subset of $\bigcup S$ and hence, will be enumerated. Consequently, by Part (a) of the inductive step proven above, \algo$({\cal T}_{init}, {\cal T}_{final}, k, I)$ returns {\em True}. Conversely, if there is an enumerated set $I$ such that \algo$({\cal T}_{init}, {\cal T}_{final}, k, I)$ returns {\em True}, then by Part (a) of the inductive step, there is a sequence of at most $k$ flips that transforms ${\cal T}_{init}$ into ${\cal T}_{final}$. Therefore, \algos$({\cal T}_{init}, {\cal T}_{final}, k, S)$ returns {\em True} if and only if $\dist({\cal T}_{init},{\cal T}_{final}) \leq k$.

\end{enumerate}
\end{enumerate}
This completes the proof.
\end{proof}

The following lemma bounds the number of leaves in the search trees of \algo$({\cal T}_{init}, {\cal T}_{final}, k, I)$ and \algos$({\cal T}_{init}, {\cal T}_{final}, k, S)$.

\begin{lemma}\label{lem:is}
Let $n = \phi({\cal T}_{init})$. Let $L_I(n,k)$ be the number of leaves in the search tree of \algo$({\cal T}_{init}, {\cal T}_{final}, k, I)$. Let $L_S(n,k,s)$ be the number of leaves in the search tree of \algos$({\cal T}_{init}, {\cal T}_{final}, k, S)$, where $s = |S|$. Then $L_I(n,k) \leq 3^{2(k-n)}$ and $L_S(n,k,s) \leq 3^{s+2(k-n)}$.
\end{lemma}
\begin{proof}
The proof is by mutual induction on $k$. For the base case when $k=0$, we have $n=0$ and both search trees have only 1 leaf. The statements are true.

Based on the inductive hypothesis, the inductive step is proven in two parts.
\begin{enumerate}[(a)]
    \item First consider \algo$({\cal T}_{init}, {\cal T}_{final}, k, I)$, which flips $|I|$ edges and create a set $S$ of $2|I|$ edge-pairs. It then calls \algos$(\overline{\cal T}_{init}, {\cal T}_{final}, k-|I|, S)$ where $\phi(\overline{\cal T}_{init}) = \phi({\cal T}_{init}) = n$. By the inductive hypothesis, the number of leaves in  \algos$(\overline{\cal T}_{init}, {\cal T}_{final}, k-|I|, S)$ is at most $3^{|S|+2(k-|I|-n)} = 3^{2|I|+2(k-|I|-n)} = 3^{2(k-n)}$. Since there is no branching in \algo, we have $L_I(n,k) \leq 3^{2(k-n)}$ as required.

\item Now consider \algos$({\cal T}_{init}, {\cal T}_{final}, k, S)$ in two cases:
\begin{enumerate}[(i)]
    \item If there is no free-diagonal in ${\cal T}_{init}$, then \algos$({\cal T}_{init}, {\cal T}_{final}, k, S)$ branches into up to 3 subtrees for each pair in $S$, corresponding to three possible choices that include at most one edge in the pair. Therefore, at most $3^{|S|}$ subtrees are created after all edge-pairs in $S$ are branched on and each subtree is a call to \algo$({\cal T}_{init}, {\cal T}_{final}, k, I)$. By Part (a) of the inductive step proven above, \algo$({\cal T}_{init}, {\cal T}_{final}, k, I)$ has at most $3^{2(k-n)}$ leaves. Therefore, the total number of leaves in the search tree of \algos$({\cal T}_{init}, {\cal T}_{final}, k, S)$ is at most $3^{|S|}3^{2(k-n)} = 3^{s+2(k-n)}$, as required.
    
    \item Suppose that a free-diagonal $e$ is flipped in Step 1 of \algos~to create a new diagonal $\overline{e} \in {\cal T}_{final}$. The algorithm partitions the triangulations $\overline{\cal T}_{init}$ and ${\cal T}_{final}$ along $\overline{e}$ into $\{\overline{\cal T}_{init}^1,  \overline{\cal T}_{init}^2\}$ and $\{{\cal T}_{final}^1,{\cal T}_{final}^2\}$, respectively. This creates two smaller instances $(\overline{\cal T}_{init}^1, {\cal T}_{final}^1)$ and $(\overline{\cal T}_{init}^2,~ {\cal T}_{final}^2)$. We may assume both are non-trivial because, by Lemma~\ref{lem:trivial-linear}, trivial instances are solvable in linear time. Let $n_1 = \phi(\overline{\cal T}_{init}^1)$. The algorithm calls \algos$(\overline{\cal T}_{init}^1, {\cal T}_{final}^1, \hat{k}_1, S_1)$ for $\hat{k}_1 = n_1+1, \ldots, k-2-n_2$, until $\hat{k}_1 = k_1$, where $k_1 = \dist(\overline{\cal T}_{init}^1, {\cal T}_{final}^1)$. If no such $k_1$ is found, then the search tree terminates and we will show in the end that the statement of the lemma is true. Since $\hat{k}_1 < k$, by the inductive hypothesis, \algos$(\overline{\cal T}_{init}^1, {\cal T}_{final}^1, \hat{k}_1, S_1)$ has at most $3^{|S_1|+2(\hat{k}_1 - n_1)}$ leaves. This means for $\hat{k}_1 = n_1+1, \ldots, k_1$, the subtrees of \algos$(\overline{\cal T}_{init}^1, {\cal T}_{final}^1, \hat{k}_1, S_1)$ has at most $3^{|S_1|+2}$, $3^{|S_1|+4}$, $\ldots$, $3^{|S_1|+2(k_1 - n_1)}$ leaves, respectively. Observing that this is a geometric sequence with a common ratio of 9, their sum is at most $\frac{9}{8}\cdot3^{|S_1|+2(k_1 - n_1)}$. 
    
    Finally, the algorithm calls \algos$(\overline{\cal T}_{init}^2,~ {\cal T}_{final}^2,~ k -1- k_1,~ S_2)$, which by the inductive hypothesis has at most $3^{|S_2| + 2(k -1- k_1 - n_2)}$ leaves.

    Thus, in total, the number of leaves in the search tree of \algos$({\cal T}_{init}, {\cal T}_{final}, k, S)$ is at most $\frac{9}{8}\cdot3^{|S_1|+2(k_1 - n_1)} + 3^{|S_2| + 2(k -1- k_1 - n_2)}$. Let $x = 3^{|S_1|+2(k_1 - n_1)}$ and $y = 3^{|S_2| + 2(k -1- k_1 - n_2)}$. Since both instances $(\overline{\cal T}_{init}^1, {\cal T}_{final}^1)$ and $(\overline{\cal T}_{init}^2,~ {\cal T}_{final}^2)$ are non-trivial, we have $k_1 - n_1 \geq 1$ and $k -1- k_1 - n_2 \geq 1$, and hence $x,y \geq 9$. Therefore, the number of leaves in the search tree of \algos$({\cal T}_{init}, {\cal T}_{final}, k, S)$ is at most
    \begin{align} \label{eqn}
    \frac{9}{8}\cdot3^{|S_1|+2(k_1 - n_1)} + 3^{|S_2| + 2(k -1- k_1 - n_2)} &= \frac{9}{8}x + y\\
    &=(\frac{27}{8y} + \frac{3}{x}) \cdot \frac{xy}{3}\\
    &\leq (\frac{27}{8\cdot 9} + \frac{3}{9}) \cdot \frac{xy}{3}\\
    &\leq \frac{xy}{3} \\
    & = \frac{3^{|S_1|+2(k_1 - n_1)} \cdot 3^{|S_2| + 2(k -1- k_1 - n_2)}}{3}\\
    & = 3^{|S_1|+|S_2|-1+2(k -1- n_1 - n_2)}.
    \end{align}
We have $|S_1|+|S_2| \leq |S|+1$ because when a free-diagonal $e$ is flipped, at least one edge-pair in $S$ that contains $e$ is removed (by Lemma~\ref{lem:new-free} and Claim~\ref{cla:all}, $e$ must belong to at least one edge pair in $S$) and two new edge-pairs are added to $S$. We also have $n_1 + n_2 = n-1$ because after the $\overline{\cal T}_{init}$ is partitioned along the diagonal $\overline{e}$, the total number of diagonals is reduced by 1. 
    
    Therefore, the total number of leaves in the search tree of \algos$({\cal T}_{init}, {\cal T}_{final}, k, S)$ is $L_S(n,k,s) \leq 3^{|S_1|+|S_2|-1+2(k -1- n_1 - n_2)} \leq 3^{|S|}3^{2(k-n)} = 3^{s+2(k-n)},$ as required.

    Now consider the special case when the search for $k_1$ is not successful in Step 1.5. In this case, \algos$(\overline{\cal T}_{init}^1, {\cal T}_{final}^1, \hat{k}_1, S_1)$ is called for $\hat{k}_1 = n_1+1, \ldots, k-2-n_2$. For the same reason as above, the total number of leaves is a sum of a geometric sequence and is bounded by $\frac{9}{8}\cdot3^{|S_1|+2(k-2-n_2 - n_1)} = \frac{9}{8}\cdot3^{|S_1|+2(k-n)} \leq \frac{9}{8}\cdot3^{|S_1|+2(k-1-n)} \leq 3^{|S_1|+2(k-n)} \leq 3^{s+2(k-n)}$, as required. Note that in this case, the search tree terminates and there is no need to call \algos$(\overline{\cal T}_{init}^2,~ {\cal T}_{final}^2,~ k -1- k_1,~ S_2)$.

\end{enumerate}
\end{enumerate}
This completes the proof.
\end{proof}

When the algorithm initially starts, the set of source nodes may be any subsets $I$ of independent diagonals in ${\cal T}_{init}$. We will prove in Lemma~\ref{lem:iterate} that there are at most $F_{n+1}$ such subsets, where $F_n$ is the $n$-th Fibonacci number, and they can be enumerated using polynomial space and in time $\Oh(n)$ each. 

\begin{lemma}\label{lem:iterate}
Let ${\cal T}$ be a triangulation of a convex polygon and $\phi({\cal T}) = n$. The number of subsets of independent diagonals in ${\cal T}$ is at most $F_{n+1}$, the $(n+1)$-th Fibonacci number. Furthermore, all such subsets can be iterated in time $\Oh(n)$ each using polynomial space.
\end{lemma}
\begin{proof}
First, observe that the set of subsets of independent diagonals in ${\cal T}$ is in bijection with the set of matchings in the binary tree $T$ corresponding to ${\cal T}$. Each triangle in ${\cal T}$ corresponds to a node in $T$ and each diagonal shared by two triangles in ${\cal T}$ corresponds to an edge between the two nodes in $T$ representing the two triangles. A set of independent diagonals in ${\cal T}$ corresponds to a subset of edges in $T$ that do not share any endpoint, referred to as a {\em matching} (we consider an empty set to be a matching). By Lemma~\ref{lem:matching}, the number of matchings in $T$ is at most $F_{n+1}$ and hence the number of subsets of independent diagonals in ${\cal T}$ is at most $F_{n+1}$. 

To iterate all such subsets $I$, consider all diagonals in an arbitrarily fixed order. For each diagonal $e$, if any of its neighbors has already been included in $I$, do not include $e$; otherwise, branch on $e$ to include or exclude $e$ in $I$. When all diagonals have been branched on, $I$ is a subset of independent diagonals in ${\cal T}$. Since each leaf of this branching tree corresponds to a unique subset of independent diagonals in ${\cal T}$ and the depth of the branching tree is $n$, the running time is $\Oh(n)$ for each subset. The iteration uses polynomial space because only a subset of diagonals in ${\cal T}$ is maintained at each step of the branching.
\end{proof}

Finally, we have the correctness and complexity of our algorithm.
\begin{theorem}
The algorithm \alg$({\cal T}_{init}, {\cal T}_{final}, k)$ runs in time $\Oh(3.82^k)$ using polynomial space and returns {\em True} if and only if there is a sequence of at most $k$ flips that transforms ${\cal T}_{init}$ into ${\cal T}_{final}$. 
\end{theorem}
\begin{proof}
By Lemma~\ref{lem:iterate}, \alg$({\cal T}_{init}, {\cal T}_{final}, k)$ branches into at most $F_{n+1}$ subsets of independent diagonals. For each such subset $I$, the function \algo$({\cal T}_{init}, {\cal T}_{final}, k, I)$ is called, which has at most $3^{2(k-n)}$ leaves by Lemma~\ref{lem:is}. Therefore, the search tree of \alg$({\cal T}_{init}, {\cal T}_{final}, k)$ has at most $F_{n+1}3^{2(k-n)}$ leaves, which means there are at most $F_{n+1}3^{2(k-n)}$ root to leaf paths in the search tree. 

Along each root-to-leaf path, the algorithm does the following: (a) enumerates the initial independent set $I$ in time $\Oh(n)$ when charged to each path; (b) performs at most $k$ flips that take time $\Oh(1)$ each; (c) performs at most $n$ partitions that take time $\Oh(n)$ each; (d) perform at most $k$ rounds of branching, where each round takes time $\Oh(k)$ when charged to each end-branch. Therefore, the time spent on each root-to-leaf path is $\Oh(k^2+n^2) = \Oh(n^2)$. Since $\Oh(n^2F_{n+1}) = \Oh(1.618^n)$, the overall running time is $\Oh(n^2F_{n+1}3^{2(k-n)}) = \Oh(1.618^n3^{2(k-n)}) = \Oh\left(9^k\cdot\left(\frac{1.618}{9}\right)^n\right)$.

Since $k/2 \leq n \leq k$, the overall running time $\Oh\left(9^k\cdot\left(\frac{1.618}{9}\right)^n\right)$ is maximized when $n = k/2$. Therefore, the total running time of the algorithm is $\Oh\left(9^k\cdot\left(\frac{1.618}{9}\right)^n\right) = \Oh\left(9^k\cdot\left(\frac{1.618}{9}\right)^{k/2}\right)= \Oh(3.82^k).$

\alg$({\cal T}_{init}, {\cal T}_{final}, k)$ enumerates all subsets $I$ of independent diagonals as the initial set of source nodes and calls \algo$({\cal T}_{init}, {\cal T}_{final}, k, I)$. If there is a sequence $F$ of at most $k$ flips that transforms ${\cal T}_{init}$ into ${\cal T}_{final}$, then the set $I$ of source nodes in ${\cal D}_F$ is enumerated and hence by Lemma~\ref{thm:correct}, \algo$({\cal T}_{init}, {\cal T}_{final}, k, I)$ returns {\em True}. Conversely, if for a certain set $I$ of independent diagonals enumerated by the algorithm, \algo$({\cal T}_{init}, {\cal T}_{final}, k, I)$ returns true, then there is valid a sequence of at most $k$ flips that transforms ${\cal T}_{init}$ into ${\cal T}_{final}$. Therefore, the algorithm is correct.

Finally, the algorithm uses polynomial space because every step of it, including the iteration of the initial independent sets (Lemma~\ref{lem:iterate}), uses polynomial space. 

This proves the correctness and complexity of our algorithm.

\end{proof}

\section{Concluding remarks}
Both {\sc Convex Flip Distance} and {\sc General Flip Distance} are important problems. The current paper presents a simple $\FPT$~algorithm for {\sc Convex Flip Distance} that runs in time $\Oh(3.82^k)$ and uses polynomial space, significantly improving the previous best $\FPT$~algorithm the problem, which runs in time $\Oh(n + k \cdot 32^k)$ and is the same $\FPT$~algorithm for {\sc General Flip Distance}~\cite{feng_improved}. 

Our algorithm takes advantage of the structural properties of {\sc Convex Flip Distance} regarding the common diagonals and the free-diagonals. However, the general approach of our algorithm, namely finding a topological sort of the DAG ${\cal D}_F$ by repeatedly removing the source nodes, seems applicable to {\sc General Flip Distance}. It remains to be seen if can be used to derive an improved algorithm for {\sc General Flip Distance}. 

The recent progress on both {\sc Convex Flip Distance} and {\sc General Flip Distance}, including~\cite{fd17, feng_improved} and this work, all rely on the DAG that models dependency relation among the flips. More research along this line will likely produce further improved algorithms. However, deciding whether the {\sc Convex Flip Distance} problem is \NP-hard remains a challenging open problem and may require new insights into the structural properties of the problem.

\section*{Declarations}
The algorithm presented in this paper has been implemented in C++ and is available in the GitHub repository, https://github.com/syccxcc/FlipDistance.


\bibliography{article}

\end{document}